\documentclass[twocolumn,prb,showpacs]{revtex4}

\usepackage{graphicx}
\usepackage{dcolumn}
\usepackage{bm}

\begin{document}

\title{Anomalous behaviors of the charge and spin degrees 
of freedom \\
in the CuO double chains of PrBa$_2$Cu$_4$O$_8$}

\author{S.~Nishimoto}
\email{satoshi.nishimoto@physik.uni-marburg.de}
\affiliation{Philipps-Universit\"at Marburg,
Fachbereich Physik, D-35032 Marburg, Germany} 

\author{Y.~Ohta}
\email{ohta@science.s.chiba-u.ac.jp}
\affiliation{Department of Physics, Chiba University,
Chiba 263-8522, Japan}

\date{\today}

\begin{abstract}
The density-matrix renormalization-group method is used to 
study the electronic states of a two-chain Hubbard model 
for CuO double chains of PrBa$_2$Cu$_4$O$_8$.  
We show that the model at quarter filling has the charge 
ordered phases with stripe-type and in-line--type patterns 
in the parameter space, and in-between, there appears a 
wide region of vanishing charge gap; the latter phase is 
characteristic of either Tomonaga-Luttinger liquid or a 
metallic state with a spin gap.  
We argue that the low-energy electronic state of the CuO 
double chains of PrBa$_2$Cu$_4$O$_8$ should be in the 
metallic state with a possibly small spin gap.  
\end{abstract}

\maketitle


\section{INTRODUCTION}

Geometrical frustration in the charge degrees of freedom 
of the CuO double chains of PrBa$_2$Cu$_4$O$_8$ has 
provided an interesting opportunity for studying anomalous 
metallic states realized in quasi-one-dimensional (1D) 
electronic conductors.  Experimentally, it has been 
reported~\cite{Hor00,Hor02,Hussey} that the system shows 
metallic conductivity along the chain direction down to 
2K and that it shows insulating conductivity along the 
directions perpendicular to the chain above $\sim$140 K.  
The system may thus be regarded as a quasi-1D correlated 
conductor except at low temperatures.  Below $\sim$140 K, 
the conductivity of the system shows a 2D behavior within 
the crystallographic $a$-$b$ plane,~\cite{Hor02} indicating 
that the coupling between the chains is not negligible.  

Anomalous behaviors in PrBa$_2$Cu$_4$O$_8$ have been 
observed in particular in its charge degrees of freedom.  
A nuclear-quadrupole-resonance (NQR) experiment~\cite{Fuji} 
showed that the temperature dependence of the spin-lattice 
($1/T_1$) and spin-spin ($1/T_2$) relaxations is anomalous, 
indicating the presence of very slow fluctuations of charge 
carriers.  A polarized optical spectrum for ${\bf E}\parallel 
a$ (electric field ${\bf E}$ parallel to the $a$ axis) 
demonstrated that the CuO$_2$ planes of this material 
maintain the charge-transfer (CT) insulating state with a 
CT gap of 1.4 eV,~\cite{Tak00,Feh93} which is typical of 
that of cuprates, but the spectrum for ${\bf E}\parallel b$ 
(chain direction) exhibits a broad peak structure at 
$\omega \gtrsim 40$ meV with a very small but nonzero Drude 
weight ($\sim$2\% of the total weight).~\cite{Tak00}  
Tomonaga-Luttinger liquid (TLL) behavior was also suggested; 
from the power-law dependence dominating in the higher-energy 
part of the optical conductivity, the exponent $K_\rho$ was 
estimated to be $\sim 0.24$.~\cite{Tak00}  Moreover, an 
angle-resolved photoemission spectroscopy (ARPES) study of 
Zn-doped PrBa$_2$Cu$_4$O$_8$ showed a power-law behavior of 
the spectral function near the Fermi level, which is indicative 
of a TLL state, and a value $K_\rho \sim 0.24$ was 
estimated.~\cite{Miz00,Miz02}  Such a small value of $K_\rho$ 
suggests that the long-range Coulomb interaction between 
charge carriers should be very strong.  It was also reported 
that the system of the CuO double chains is near quarter 
filling of holes.~\cite{Tak00,Miz00}  A good opportunity 
has thus been provided for studying the anomalous charge 
dynamics of strongly correlated 1D electron systems such 
as charge ordering (CO), charge fluctuation, and charge 
frustration.  

A few theoretical studies have so far been made on the CuO 
double chains of PrBa$_2$Cu$_4$O$_8$.  Using Lanczos 
exact-diagonalization technique on small clusters, Seo and 
Ogata~\cite{Seo01} studied similar models and proposed 
possibility of the appearance of a metallic phase due to 
geometrical frustration; the system sizes used were 
however very small, so that whether the first charge 
excitation is gapped or gapless in the thermodynamic limit 
was not clear.  Zhuravlev {\it et al.}~\cite{Zhu97} 
investigated a spinless Fermion model at half filling, 
which corresponds to a quarter-filled extended Hubbard 
model in the limit of large on-site repulsion $U$, and 
found that a metallic state is realized due to the effect 
of frustration among Fermions when long-range repulsive 
interactions among them compete.  Amasaki 
{\it et al.}~\cite{Ama02} calculated the optical 
conductivity spectrum in the mean-field theory and showed 
that it agrees well with experiment only when the stripe-type 
CO is present; they thereby argued that the observed 
anomalous optical response may be due to the presence of 
the stripe-type fluctuations of charge carriers in the CuO 
double chains.  

Motivated by such development in the field, we here 
investigate the ground state of the extended Hubbard 
model at quarter filling, an effective model for the CuO 
double chain of PrBa$_2$Cu$_4$O$_8$.  
An isolated double-chain system is considered in this 
paper because to clarify its nature is essential before 
the `dimensional crossover' is taken into account.  
We use the density-matrix renormalization-group (DMRG) 
method~\cite{Whi93} for calculating the ground and 
low-energy excited states of the model, as well as its 
charge and spin correlation functions.  

We will show that our model in its ground state has the 
CO phases with stripe-type and in-line--type patterns 
in the parameter space, and in-between, there appears 
a wide region of vanishing charge gap; the latter 
phase is characteristic of either a TLL or a metallic 
state with a spin gap.  The spin gap remains open unless 
the filling deviates largely from a quarter.  We will 
then consider possible experimental relevance of our 
results.  We will in particular argue that the low-energy 
electronic state of the CuO double chains of 
PrBa$_2$Cu$_4$O$_8$ should be in the metallic state 
with a possibly small spin gap.  

This paper is organized as follows.  In Sec.~II, we 
introduce a single-band Hubbard model defined on a 1D 
lattice for the CuO double chain of PrBa$_2$Cu$_4$O$_8$ 
and give some numerical details of the DMRG method used.  
In Sec.~III, we present calculated results for the charge 
gap, phase diagram, spin correlations, charge correlations, 
and Luttinger parameters, and discuss anomalous properties 
of the liquid phase of our system.  
In Sec.~IV, we summarize the results obtained and discuss 
their possible experimental relevance.

\section{MODEL AND METHOD}

\begin{figure}
\includegraphics[width=6cm]{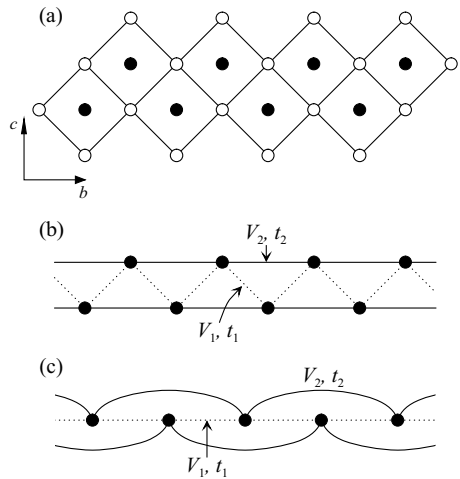}
\caption{\label{fig1} (a) Schematic representation of the 
lattice structure of the CuO double chain.  The solid 
circles represent Cu $(d_{x^2-y^2})$ orbitals and the 
open circles represent O $(p_\sigma)$ orbitals.  
(b) The corresponding two-chain Hubbard model.  
(c) Geometrically equivalent single-chain Hubbard model.}
\end{figure}

The structural element of PrBa$_2$Cu$_4$O$_8$ that we 
consider is the CuO double chain with unit cell 
Cu$_2$O$_4$,~\cite{Hor00,Hor02} the geometry of which is 
illustrated in Fig.~\ref{fig1}(a).  The relevant orbitals 
on the Cu and O sites are the $3d_{x^2-y^2}$ and 
$2p_{\sigma}$ orbitals, respectively.  From the antibonding 
band of these two orbitals, we may extract an effective 
model, i.e., a single-band extended Hubbard model for 
low-energy electronic states of the CuO double chains 
(see Fig.~\ref{fig1}(b)), where the O ions are no longer 
explicitly present.  Such reduction of dimensionality 
allows us to carry out detailed numerical studies.  One 
should also note that this two-chain Hubbard model is 
topologically equivalent to the single-chain Hubbard model 
with next-nearest-neighbor interactions as shown in 
Fig.~\ref{fig1}(c).  

The model we consider is thus described by the Hamiltonian 
\begin{eqnarray}
H = &-& t_1 \sum_{i\sigma} (c^\dagger_{i\sigma}
c_{i+1\sigma} + c^\dagger_{i+1\sigma} c_{i\sigma})\nonumber\\
&-& t_2 \sum_{i\sigma} (c^\dagger_{i\sigma}
c_{i+2\sigma} + c^\dagger_{i+2\sigma} c_{i\sigma})\nonumber\\
&+& U \sum_i n_{i\uparrow}n_{i\downarrow}\nonumber\\
&+& V_1 \sum_{i\sigma\sigma^\prime} \left(n_{i\sigma}-\frac{1}{2}\right)
\left(n_{i+1\sigma^\prime}-\frac{1}{2}\right)\nonumber\\
&+& V_2 \sum_{i\sigma\sigma^\prime} \left(n_{i\sigma}-\frac{1}{2}\right)
\left(n_{i+2\sigma^\prime}-\frac{1}{2}\right)
\label{hamiltonian}
\end{eqnarray}
where we use the notations for the 1D chain defined in 
Fig.~\ref{fig1}(c).  $c^\dagger_{i\sigma}$ ($c_{i\sigma}$) 
is the creation (annihilation) operator of a hole with 
spin $\sigma$ $(=\uparrow,\downarrow)$ at site $i$ and 
$n_{i\sigma}=c^\dagger_{i\sigma}c_{i\sigma}$ is the 
number operator.  $t_1$ and $V_1$ are the nearest-neighbor 
hopping integral and Coulombic repulsion, respectively, 
which are for the zigzag bonds connecting two chains of 
the two-chain model (see Fig.~\ref{fig1}(b)).  $t_2$ and 
$V_2$ are those for the next-nearest-neighbor bonds, which 
are for the chain bonds of the two-chain model.  
We assume $t_1 \ll t_2$ because of the bonding angle 
minimizing the overlap the Cu $3d_{x^2-y^2}$ and O 
$2p_{\sigma}$ wave functions for the hopping parameter 
$t_1$; in the following, we use the value $t_1/t_2=0.1$ 
unless otherwise stated, by confirming that the sign of 
$t_1$ does not change the results.  We assume a value of 
the on-site Coulomb interaction $U$ as $U/t_2=20$, which 
is somewhat larger than a typical value for cuprate 
materials; by using this value, however, we can investigate 
interesting features of our model with a wide range of 
$V_1$ and $V_2$.  We believe that this choice does not 
change the essential features of our results.  
The filling of holes in the CuO double chains is reported 
to be $n \sim 0.5$,~\cite{Tak00,Miz00}; we therefore 
restrict ourselves to the case at quarter filling unless 
otherwise indicated, i.e., $n=\frac{1}{2}$ corresponding 
to a density of one hole per two Cu sites.  Note that the 
repulsions $V_1$ and $V_2$ are frustrating interactions 
for the charge degrees of freedom of the system in the 
sense that two holes can choose to sit on two of the 
three sites of a triangle of the lattice in mutually 
competing energy scales $V_1$ and $V_2$ 
(see Fig.~\ref{fig1}(b)).  If we assume the intersite 
Coulomb repulsions to be inversely proportional to the 
intersite distance, we have the relation $V_1=\sqrt{2}V_2$, 
which is not far from the line $V_1=2V_2$ where the two 
repulsions $V_1$ and $V_2$ exactly compete at $t_1=t_2=0$.  

We use the DMRG method,~\cite{Whi93} a powerful numerical 
technique for a variety of 1D systems, whereby we can obtain 
very accurate ground-state energies and expectation values 
for very large finite-size systems.  We study the chains 
with up to 128 sites.  The open-end boundary conditions are 
used; when we calculate local quantities, they are actually 
the ones averaged over the chain and when we show correlation 
functions, they are the ones measured from the midpoint of 
the chain.  We use up to $m=2000$ density matrix eigenstates 
to build the DMRG basis.  One needs to keep relatively large 
values of $m$ to obtain sufficient accuracies in our system 
because the long-range hopping term increases the truncation 
error rapidly; this is in particular the case when 
$t_1\simeq 0$.  Extrapolating the DMRG ground-state energies 
calculated for different $m$ values to an energy for vanishing 
discarded weight,~\cite{Bon00} we obtain the ground state 
energy and excitation gap which are accurate to parts in 
$10^{-3}t_2$.  Although the largest source of errors in our 
calculations are finite size effects, we are able to 
extrapolate a number of gap values measured for different 
system sizes to a value for an infinite system-size chain.  

In the following, we first present calculated results for 
the charge gap in order to show that the geometrical 
frustration indeed induces the metal-insulator (MI) phase 
transition; the ground-state phase diagram is shown 
in the parameter space of the inter-site Coulomb 
interactions.  We then present calculated results for 
the density distributions, spin-spin correlation functions, 
charge velocities, and the density-density correlation 
functions, in order to discuss possible anomalous 
behaviors of the metallic phase.  

\begin{figure}
\includegraphics[width=6cm]{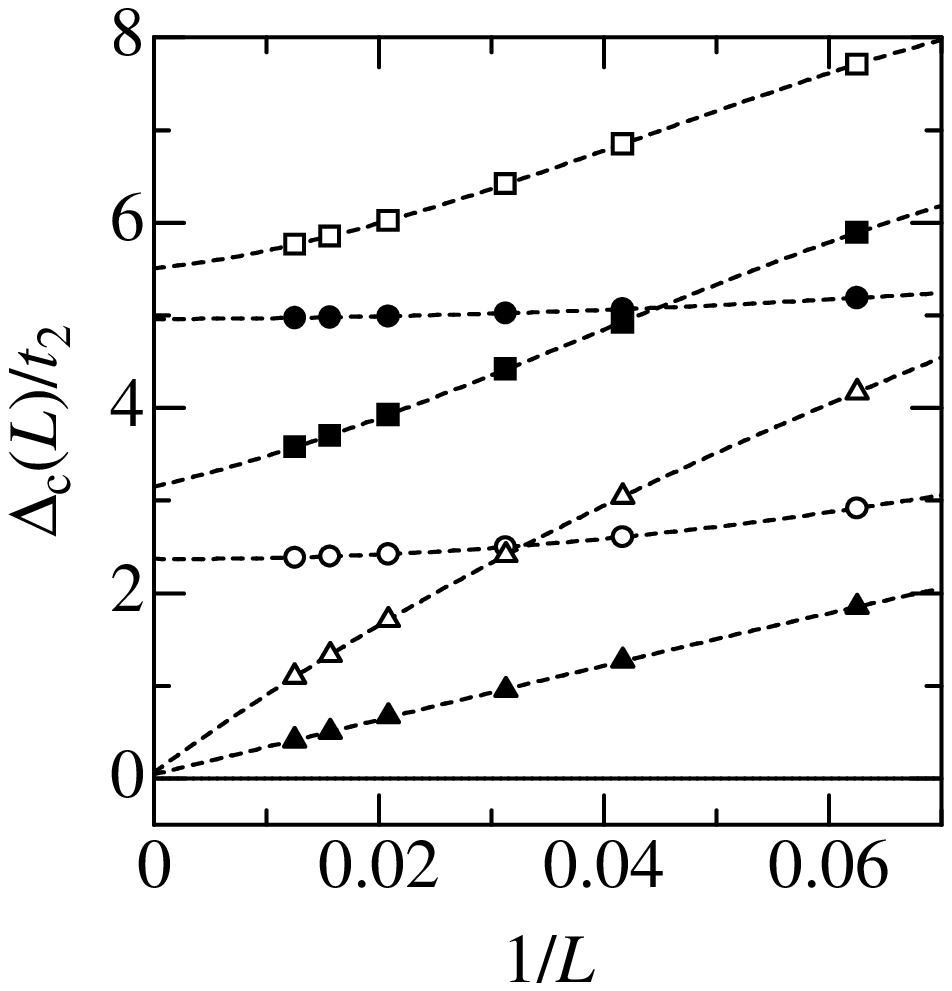}
\includegraphics[width=6cm]{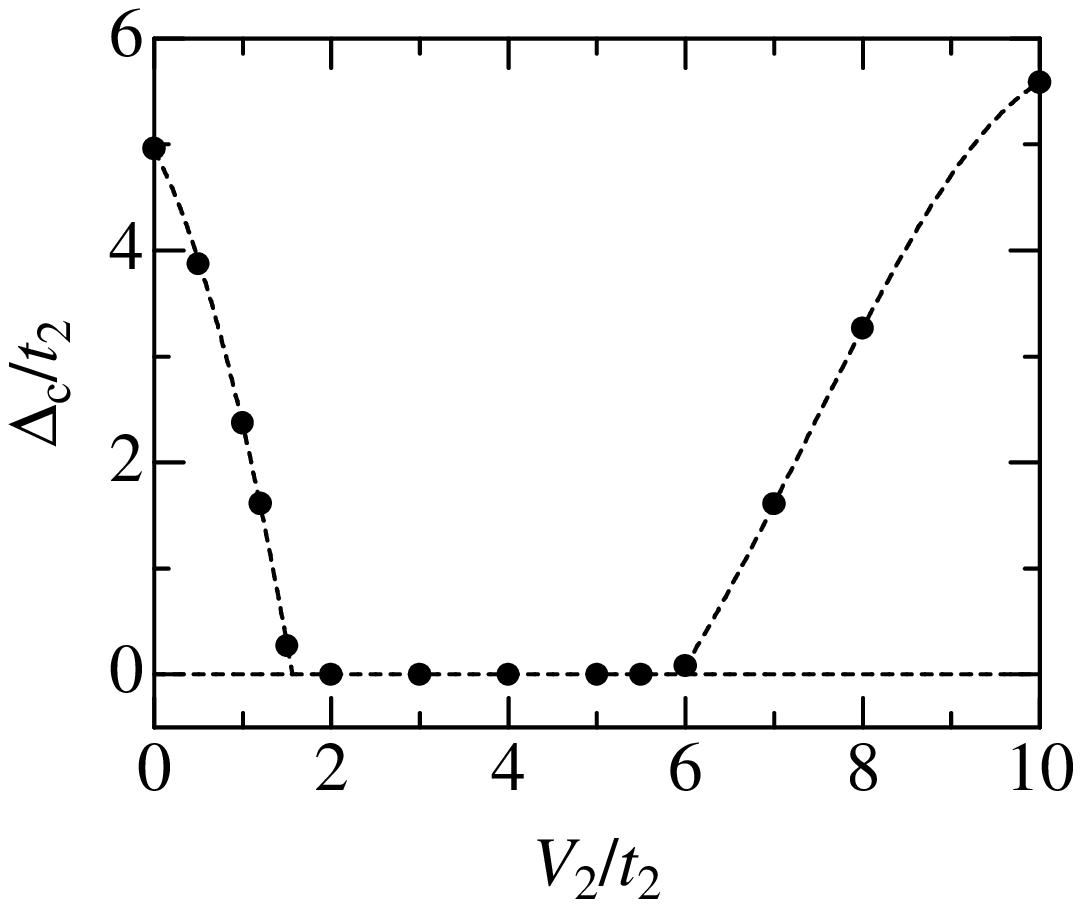}
\caption{\label{fig2} 
Charge gap in the extended Hubbard model 
[Eq.~(\ref{hamiltonian})] at quarter filling for 
$V_1/t_2=5$ and $t_1/t_2=0.1$.  
(a) $\Delta_c (L)/t_2$ calculated as a function 
of the inverse system size $1/L$ for $V_2/t_2=0$ 
(filled circles), 
$1$ (open circles), 
$2$ (filled triangles), 
$6$ (open triangles), 
$8$ (filled squares), and 
$10$ (open squares).  
Dashed lines are the quadratic fits in $1/L$.  
(b) $V_2/t_2$ dependence of $\Delta_c/t_2$ in the 
thermodynamic limit $L \rightarrow \infty$.  
The circles show the results obtained by 
extrapolation of the DMRG data in (a).  
The dashed lines are a guide to the eye.
}
\end{figure}

\section{RESULTS OF CALCULATION}

\subsection{Charge gap}

The charge gap of an $L$-site chain is defined by
\begin{eqnarray}
\Delta_c (L) &=& \frac{1}{2}
\big[ E_0 (N_\uparrow+1,N_\downarrow+1,L)
\nonumber\\
&+& E_0 (N_\uparrow-1,N_\downarrow-1,L) 
- 2E_0 (N_\uparrow,N_\downarrow,L) \big],
\label{chargegap}
\end{eqnarray}
where $E_0(N_\uparrow,N_\downarrow,L)$ denotes the 
ground-state energy of a chain of length $L$ with 
$N_\uparrow$ spin-up holes and $N_\downarrow$ 
spin-down holes.  
In Fig.~\ref{fig2}(a), we show calculated results 
for $\Delta_c (L)/t_2$ at $V_1/t_2=5$ for several 
$V_2/t_2$ values and system sizes $L$ up to 80 sites.  
We find that $\Delta_c (L)$ decreases monotonically 
with increasing $L$, so that we can extrapolate 
$\Delta_c (L)$ to the thermodynamic limit systematically 
by performing the least-square fit of $\Delta_c (L)$ 
to a second-order polynomial in $1/L$.  
The extrapolated results 
$\Delta_c/t_2$ $(=\Delta_c (L \rightarrow \infty)/t_2)$ 
are shown in Fig.~\ref{fig2}(b).  It is evident that 
for both small $V_2/t_2 \lesssim 1.56$ and large 
$V_2/t_2 \gtrsim 5.95$, $\Delta_c/t_2$ is finite, 
i.e., the system is insulating, and in a wide range 
of $V_2/t_2$, i.e., $1.56 \lesssim V_2/t_2 \lesssim 5.95$, 
$\Delta_c/t_2$ vanishes within the accuracy of the 
extrapolation.  
We find similar situations for other values of $V_1/t_2$ 
as well, which demonstrates that a stable metallic 
phase indeed exists between two insulating phases.  

When $t_1=t_2=0$, we readily find from a total-energy 
calculation that there are two ordered phases, of which 
the phase boundary locates on a $V_1=2V_2$ line.  The 
phase in the $V_1>2V_2$ region is characterized by an 
in-line CO, i.e., all the holes locate on one of the 
two chains of the two-chain model, whereas the phase 
in the $V_1<2V_2$ region is characterized by a 
stripe-type CO, i.e., holes locate alternately on sites 
of both of the two chains forming a stripe-like pattern 
along the chain direction.  The stripe-type CO corresponds 
to the $2k_F$ charge-density-wave (CDW) state and the 
in-line CO corresponds to the $4k_F$-CDW in the 
single-chain notation of Fig.\ref{fig1}(c).  
Note that, if we assume the intersite Coulomb repulsion 
to be inversely proportional to the intersite distance, 
we have $V_1=\sqrt{2}V_2$ and hence the stripe-type CO 
should be realized, but if we assume a faster decay 
of the intersite repulsion due to the effect of 
screening, the energies of the two phases can be more 
comparable.  

\begin{figure}
\includegraphics[width=6cm]{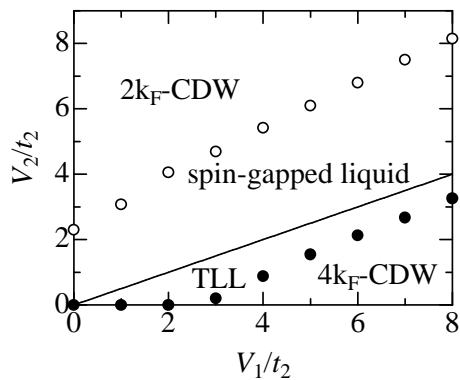}
\caption{\label{fig3} Ground-state phase diagram of 
the model [Eq.~(\ref{hamiltonian})] in the parameter space 
of the intersite Coulomb interactions.  
The values $t_1=0.1t_2$ and $U=20t_2$ are assumed.  
The open (solid) circles indicate the phase boundary 
between the $2k_F$-CDW ($4k_F$-CDW) and metallic states, 
at which the calculated charge gap $\Delta_c$ vanishes.  
The solid line corresponds to $V_2/t_2=V_1/(2t_2)$.  
Details of the metallic phase are discussed in Sec.~III E.}
\end{figure}

By making $t_1$ and $t_2$ finite to introduce quantum 
fluctuations, there appears the metallic state between 
these two insulating phases, where the two intersite 
repulsions $V_1$ and $V_2$ are mutually competing, as we 
have shown above.  The ground-state phase diagram of our 
model (see Sec.~III E for detailed discussion) therefore 
includes three phases in the parameter space as shown in 
Fig.~\ref{fig3}; 
(i) the insulating phase with the $4k_F$-CDW (or in-line 
CO) realized when $V_1$ is large, 
(ii) the insulating phase with the $2k_F$-CDW (or 
stripe-type CO) realized when $V_2$ is large, and 
(iii) the metallic phase in-between.  
These three phases are discussed further in the 
following subsections.  

\subsection{Charge ordering}

Let us first consider the three phases from the 
viewpoint of ordering of charge degrees of freedom.  
We first of all note that the CO is a state with a 
broken translational symmetry; there are two degenerate 
ground states, and in the open-end boundary conditions 
we use here, one of the two ground states is picked out 
by initial conditions of the calculation.  The CO 
state is thus directly observable in our calculations.

\begin{figure}
\includegraphics[width=6cm]{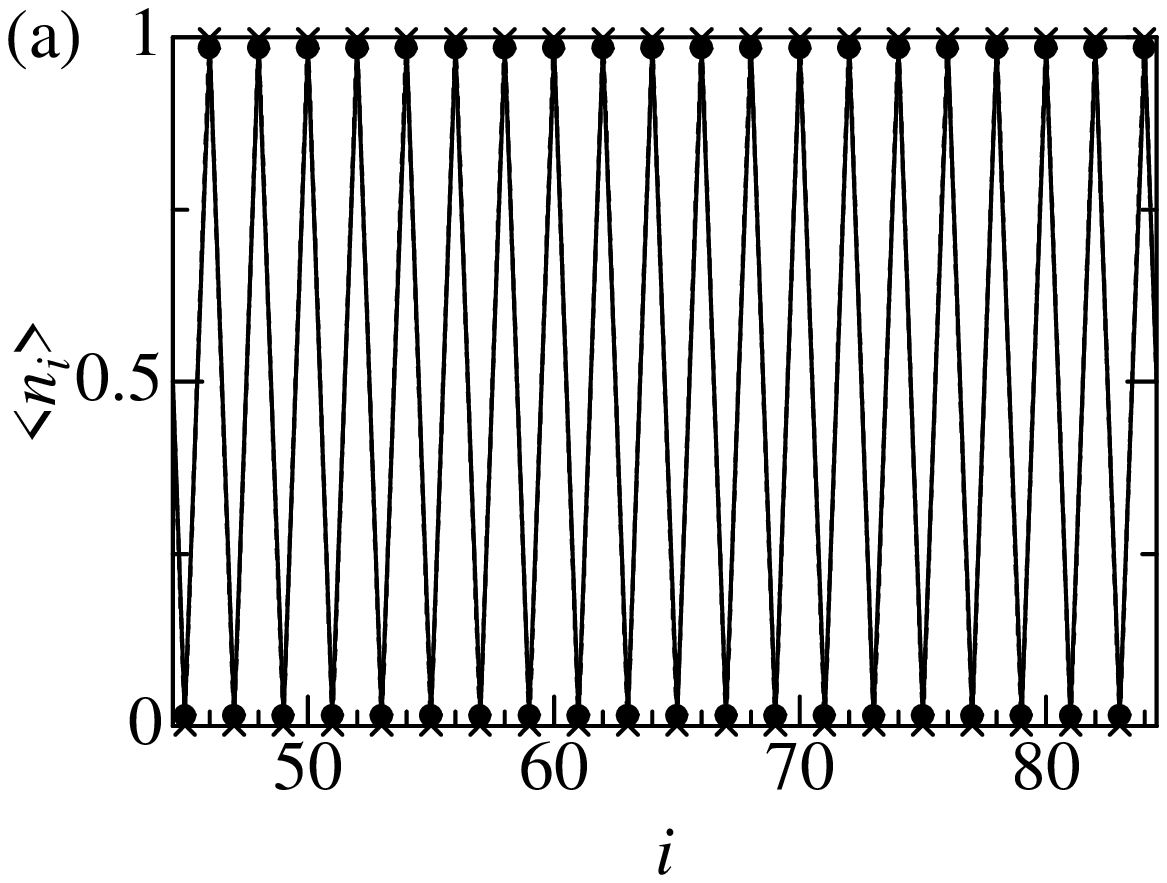}
\includegraphics[width=6cm]{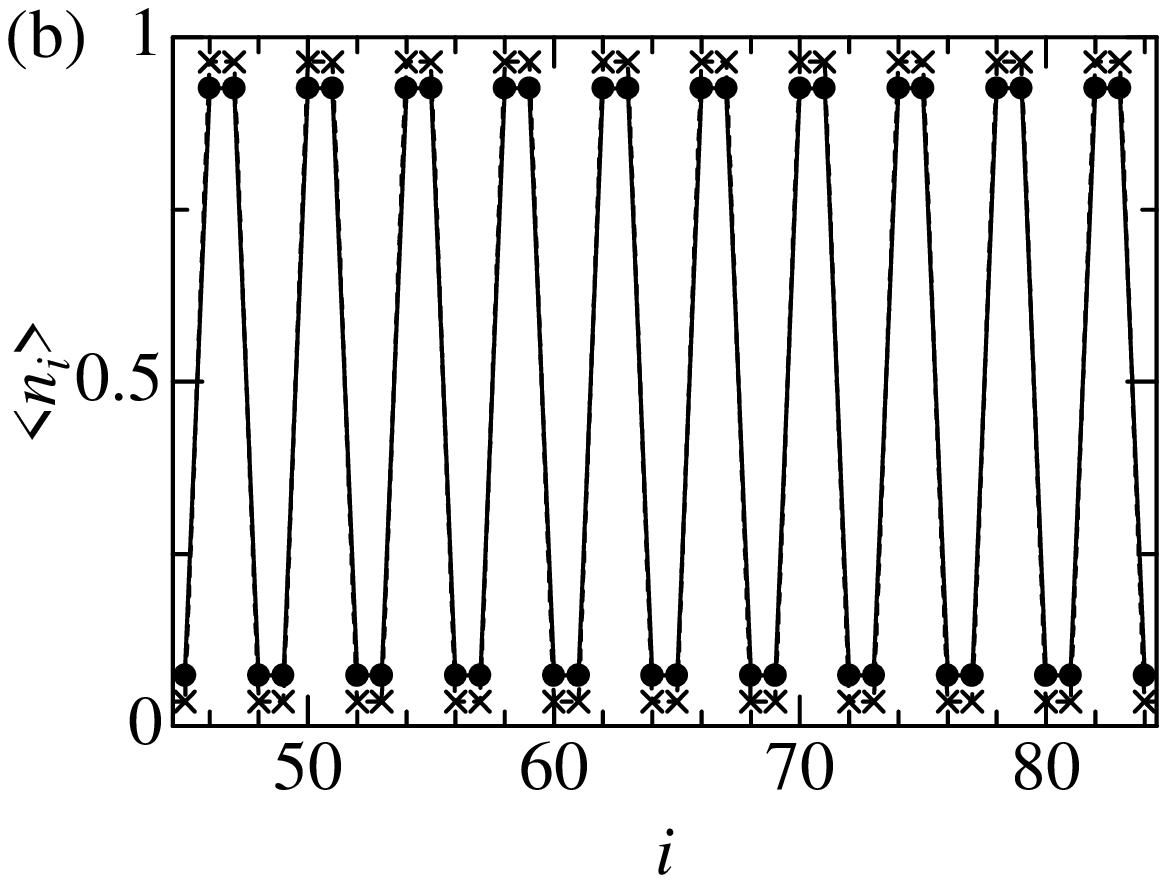}
\includegraphics[width=6cm]{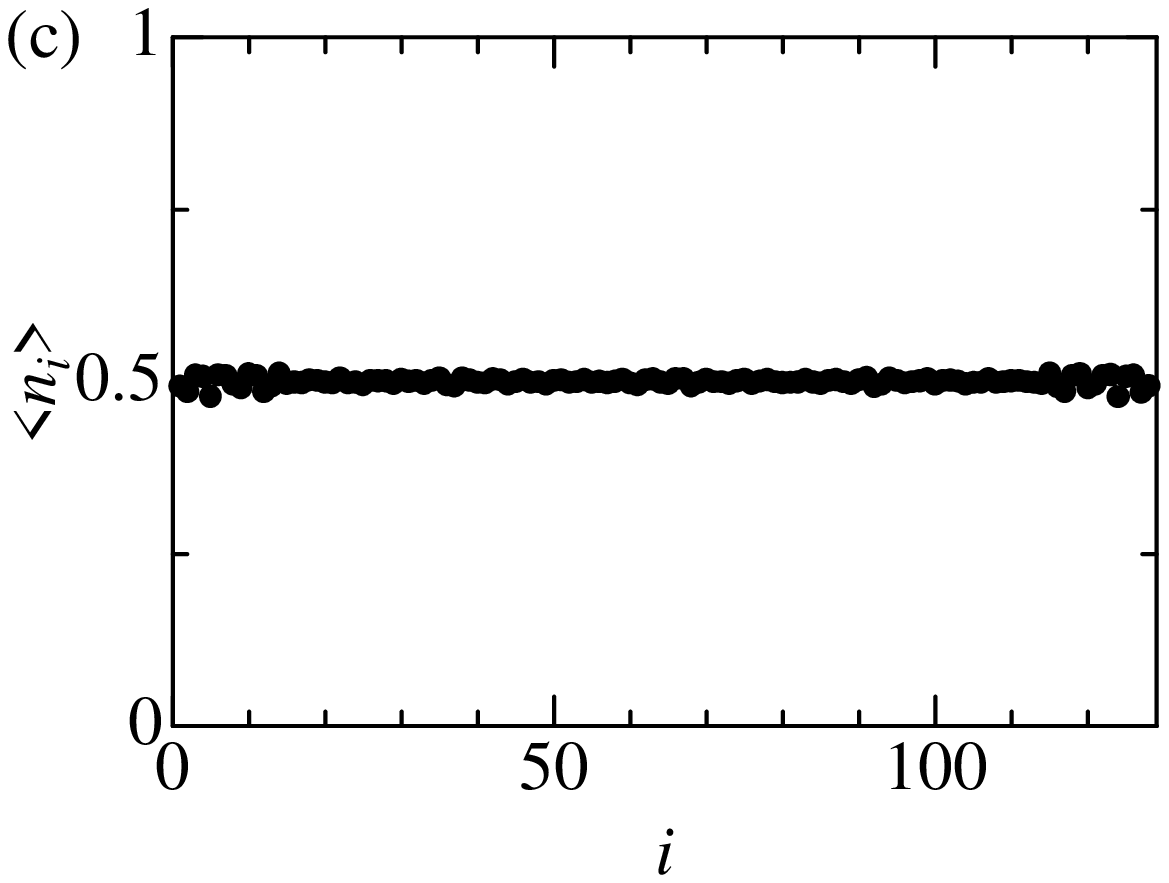}
\caption{\label{fig4} 
Calculated hole densities at site $i$ for 
(a) the $4k_F$-CDW (or in-line--type CO) at $V_2/t_2=0$ 
(crosses) and $1$ (circles), 
(b) $2k_F$-CDW (or stripe-type CO) at $V_2/t_2=6$ (circles) 
and $8$ (crosses), and 
(c) uniform metallic phase at $V_2/t_2=2$.  
We assume $V_1/t_2=5$ and use the system of $L=128$.  
Results for central 40 sites are shown in (a) and (b).  
}
\end{figure}

For small values of $V_2/t_2$ 
($\lesssim V_1/2t_2-{\cal O}(t_1)$), 
the effect of $V_1/t_2$ dominates and leads to a 
CO with wave vector $q=4k_F$ in the single-chain 
notation (which corresponds to the in-line--type 
CO in the two-chain notation); namely, the ground 
state of the system is dominated by configurations 
such as
\begin{center}
...01010101010...,
\end{center}
where ``1'' stands for a site supporting one hole 
and ``0'' stands for an empty site.  Fig.~\ref{fig4}(a) 
shows an example of the Friedel oscillation appeared 
in the density distribution $\left\langle n_i \right\rangle$ 
in the finite-size system of $L=128$.  One can see 
the $4k_F$-CDW behavior clearly.  The oscillations persist 
at the center of the chain and the difference between 
the results for $V_2/t_2=0$ and $1$ is hardly visible.  
We calculate $\left\langle n_i \right\rangle$ for several 
chain lengths and confirm that the system size with $L=128$ 
is sufficiently large and the finite-size effect is 
negligible.  We also calculate the density-density 
correlation function $N(l)$ defined as 
\begin{equation}
N(l)=\left\langle n_{i+l}n_i\right\rangle 
-\left\langle n_{i+l}\right\rangle\left\langle 
n_i\right\rangle,
\label{denscorr}
\end{equation}
with $n_i = n_{i\uparrow}+n_{i\downarrow}$ and find that 
$N(l)$ tends to a nonzero constant value as the intersite 
distance $l$ increases: e.g., 
$N(l\rightarrow\infty)\sim 10^{-7}$ at $V_1/t_2=5$ and 
$V_2/t_2=1$.  
This indicates the presence of the long-range ordered 
$4k_F$-CDW state.  

At $V_2/t_2=0$ where the $4k_F$-CDW is stabilized, most 
of the charges are located on one of the two chains of the 
two-chain model although there are some charges on the other 
chain due to quantum fluctuation by a small value of 
$t_1/t_2$; at $t_1/t_2= 0$, the charges are disproportionated 
on one of the chains completely.  
With increasing $V_2/t_2$, the charge gap decreases rapidly 
(see Fig.~\ref{fig2}(b)), but the $4k_F$-CDW correlation 
is hardly suppressed until reaching a sharp decline occurred 
around the MI transition point.  
This transition is of the second-order due to the presence 
of a small value of $t_1$; if $t_1=0$, this transition 
should be of the first-order.  
These results suggest that at $V_2/t_2<1.56$ and $V_1/t_2=5$ 
the system may be regarded as a {\em half-filled} extended 
Hubbard model since most of the charges are disproportionated.  
We note that, in the case of $t_1=1$ and $t_2=0$, the 
suppression of $4k_F$-CDW correlation by $V_2$ has been 
suggested in the weak-coupling renormalization-group 
study.~\cite{Yos01}  

For large values of $V_2/t_2$ ($\gtrsim V_1/2t_2+{\cal O}(t_1)$), 
$V_2/t_2$ enhances the $2k_F$-CDW fluctuations and induces a 
CDW with wave vector $q=2k_F$ (or a stripe-type CO in the 
two-chain notation).  The ground-state wave functions 
have dominant configurations of the type
\begin{center}
...011001100110...
\end{center}
where we use the same notations as above.  
Fig.~\ref{fig4}(b) shows an example of the Friedel 
oscillation appeared in the density distribution 
$\left\langle n_i \right\rangle$ in the finite-size 
system of $L=128$.  The $2k_F$-CDW behavior is clearly 
seen.  The calculated values of $N(l)$ tends to a 
nonzero constant as the intersite distance $l$ 
increases: e.g., $N(l\rightarrow\infty)\sim 10^{-8}$ 
at $V_1/t_2=5$ and $V_2/t_2=8$.  This indicates the 
presence of long-range ordered $2k_F$-CDW state.  

When $V_2/t_2$ is much larger than $V_1/t_2$, $2k_F$-CDW 
(or stripe-type CO) is stabilized, i.e., in the single-chain 
notation, two occupied sites and two empty sites come 
alternately.  As shown in the next subsection, two spins 
on the neighboring two sites form a spin-singlet pair.  
As $V_2/t_2$ decreases, $\Delta_c$ decreases in proportion 
to $V_2/t_2$, and its behavior is similar to that of the 
model without hybridization ($t_1=t_2=0$); i.e., 
$\Delta_c=2(V_2-V_2^{c, 2k_F})$ around the MI transition 
point $V_2^{c, 2k_F}$.  This suggests that the charges 
disproportionate almost completely as soon as the charge 
gap opens.  It can also be verified by slight reduction 
of the amplitude of the $2k_F$-CDW oscillation that occurs 
when $V_2/t_2$ decreases.  
This means that the effect of $V_1/t_2$ directly weakens 
the effect of $V_2/t_2$ and melts the $2k_F$-CDW state.  
This situation is similar to the MI transition of the 
quarter-filled extended Hubbard model.  The transition is 
of the second order for $t_2>0$; it is of the first order 
only at $t_2=0$.  

For intermediate values of $V_2/t_2$ (or around 
$V_1\sim 2V_2$), the system is in the metallic state.  
As expected, the amplitude of the Friedel oscillation 
is quite small as shown in Fig.~\ref{fig4}(c), which 
should vanish at the center of the chain when 
$L\rightarrow\infty$.  In this phase, competition of 
two incompatible fluctuations, i.e., the $2k_F$ and 
$4k_F$ (or stripe-type and in-line--type) charge 
fluctuations, prohibit the system from forming CO states.  
However, it is useful to know what is the dominant 
correlation; here, we examine the long distance behavior 
of the density-density correlations $N(l)$, in particular 
its period of oscillation.  We thus find that the $2k_F$ 
CO correlation is dominant for $2V_2 \gtrsim V_1$ and 
the $4k_F$ CO correlation is dominant for 
$2V_2 \lesssim V_1$, as will be discussed further in 
Sec.~III D.  

\begin{figure}
\includegraphics[width=6cm]{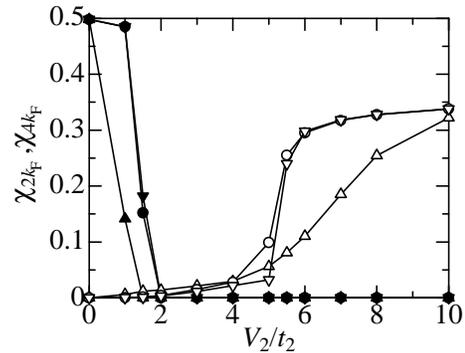}
\caption{\label{fig5} 
Order parameters $\chi_{4k_F}$ and $\chi_{2k_F}$ as a 
function of $V_2/t_2$ for various system sizes at 
$V_1/t_2=1$.  The filled (open) symbols are the results 
for $\chi_{4k_F}$ ($\chi_{2k_F}$).  The up-triangles, 
circles, and down-triangles denote the results for 
$L=16$, $48$, and $80$, respectively.}
\end{figure}

In order to explain relationship between the MI 
transition and CO formation, we introduce the order 
parameters defined as 
\begin{eqnarray}
\chi_{4k_F}=\frac{1}{L_{\it eff}}\sum_{j}e^{i4k_Fr_j}n_j 
=\frac{1}{L_{\it eff}}\sum_{j}(-1)^jn_j
\label{4kforderparameter}
\end{eqnarray}
for the $4k_F$-CDW and
\begin{eqnarray}
\chi_{2k_F}=\frac{1}{L_{\it eff}}\sum_{j}e^{i2 k_F r_j}n_j 
=\frac{1}{L_{\it eff}}\sum_{j}
\cos\left(\frac{\pi}{2}j\right)n_j,
\label{2kforderparameter}
\end{eqnarray}
for the $2k_F$-CDW.  Here, the sum is taken over central 
$L_{\it eff}$ $(=L-a\,)$ sites excluding $a/2$ sites at 
both ends of the system in order to eliminate edge 
effects; here we take $a\simeq 8-12$.  

The obtained results at $V_1/t_2=5$ as a function of 
$V_2/t_2$ are shown in Fig.~\ref{fig5} for a number 
of system sizes.  We find that as the system size 
increases the slopes of the curves around the critical 
points become steeper and that the result at $L=48$ is 
nearly equal to the result at $L=80$.  Thus, the 
situation in the thermodynamic limit may well be 
expected except near the critical point where the 
size effect is not negligible.  Let us focus on the 
$2k_F$-CDW transition at $V_2/t_2\simeq 5.95$.  
Then, $\chi_{2k_F}$ shows a sharp development around 
$V_2/t_2 \approx 5$, which is rather smaller than the 
value of $V_2^{c,2k_F}/t_2$ shown in Fig.~\ref{fig2}(b).  
However, in thermodynamic limit, this turning point 
should agree with $V_2^{c,2k_F}/t_2$, and the transition 
seems to be still continuous.  We note that the transition 
becomes sharper for larger $V_1/t_2$ (or smaller $t_2$) 
values and is of the first-order in the limit 
$V_2\to\infty$ (or $t_2\to 0$).  
A similar interpretation can also be given to the 
$4k_F$-CDW transition.  

\subsection{Spin correlations}

\begin{figure}
\includegraphics[width=6cm]{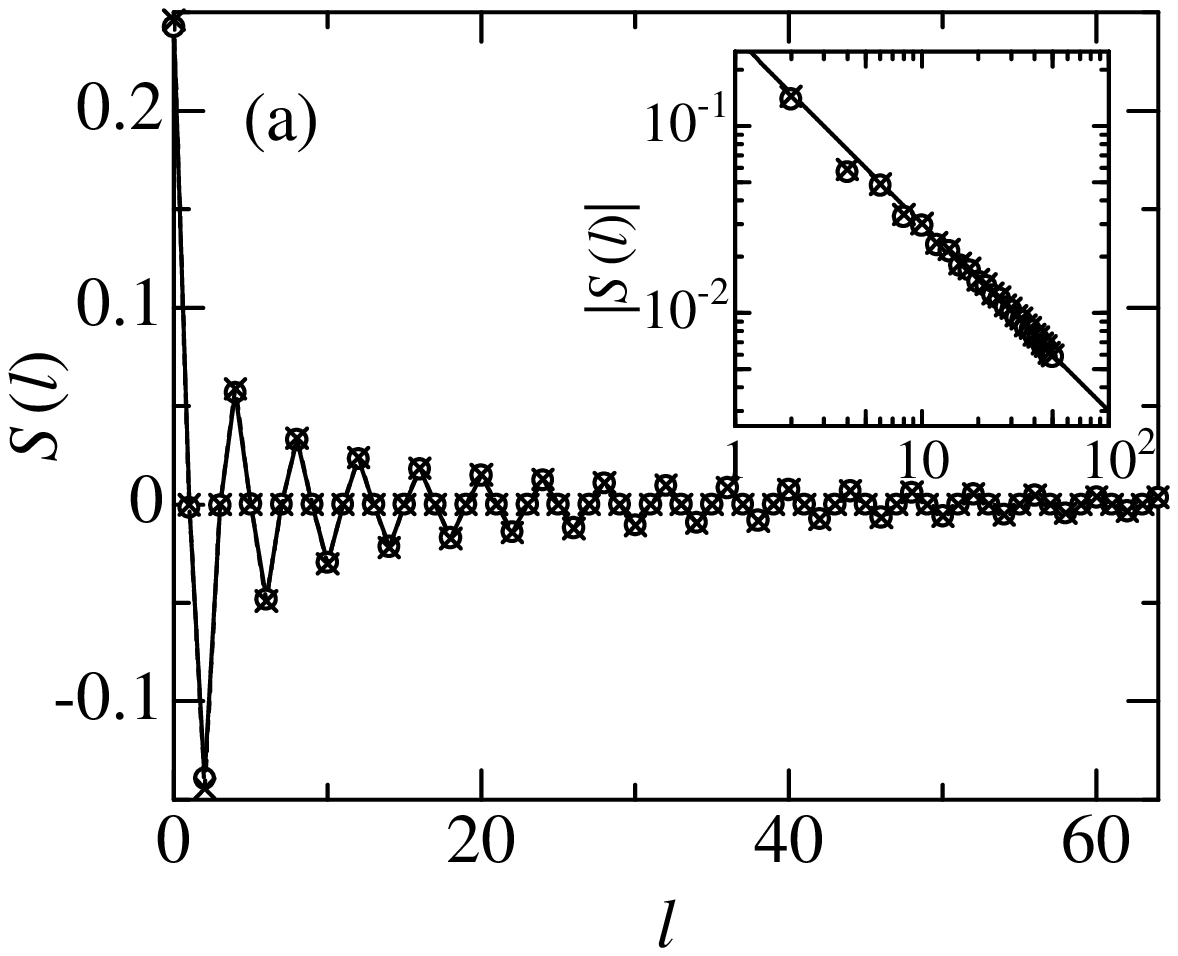}
\includegraphics[width=6cm]{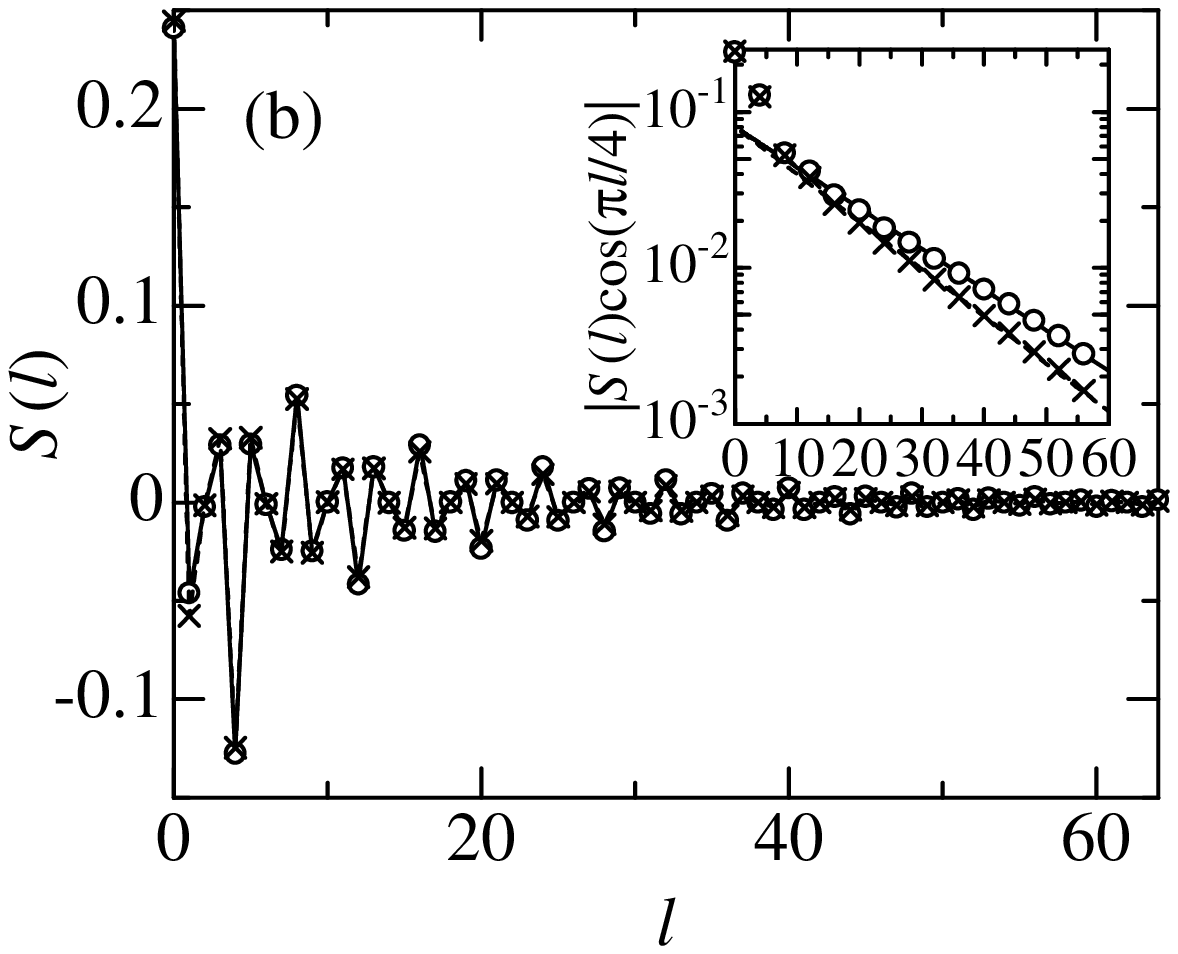}
\caption{\label{fig6} (a) Spin-spin correlation 
function $S(l)$ as defined in Eq.~(\ref{spincorr}) 
at $V_2/t_2=0$ (crosses) and $1$ (circles) with 
$V_1/t_2=5$.  The distance $l$ is measured along the 
$t_1$-chain in units of the interval between two sites.  
The log-log plot in the inset shows that the power-law 
decay of correlations.  (b) $S(l)$ at $V_2/t_2=8$ 
(circles) and $10$ (crosses) with $V_1/t_2=5$.  The 
semilog plot in the inset shows the exponential 
decay of correlations.}
\end{figure}

In order to study the spin degrees of freedom of the 
system, we now calculate the spin-spin correlation 
function defined by
\begin{equation}
S(l) = \left\langle S^z_i S^z_{i+l} \right\rangle, 
\label{spincorr}
\end{equation}
with $S^z_i = (n_{i\uparrow}-n_{i\downarrow})/2$.  

Let us first discuss the $4k_F$-CDW phase (or in-line 
CO phase in the two-chain notation).  Fig.~\ref{fig6}(a) 
shows the spin-spin correlation functions calculated 
at $V_2/t_2=0$ and $1$ with $V_1/t_2=5$ for the system 
of $L=128$.  The site $i$ in Eq.~(\ref{spincorr}) is 
fixed at the middle of the chain.  One can clearly find 
the antiferromagnetic correlations between neighboring 
spins localized on one of the two-chains of the 
two-chain model, i.e., formation of the $2k_F$ 
spin-density-wave (SDW).  One also notes that, because 
the charge fluctuations through the zigzag bonds 
connecting two chains are nearly zero due to a very 
small value of $t_1/t_2$, no spin correlations between 
two chains of the two-chain model are seen.  As shown 
in the inset of Fig.~\ref{fig6}(a), the spin-spin 
correlations decay as a power law, similar to those 
for the 1D Heisenberg chain, which depends very little 
on $V_2/t_2$ unless the $4k_F$-CDW state is broken.  
This suggests that the system may be well described as 
the 1D Heisenberg chain combined with an empty chain, 
as the in-line CO state may suggest.  

Now let us turn to the correlations in the $2k_F$-CDW phase 
(or the stripe-type CO phase in the two-chain notation).  
Fig.~\ref{fig6}(b) show the spin-spin correlation functions 
for $V_2/t_2=8$ and $10$ at $V_1/t_2=5$.  The behavior of 
$S(l)$ for $l=4n$ ($n=0, 1, 2, \cdots$) shows the 
antiferromagnetic correlations.  This corresponds to the 
$2k_F$-SDW state of one of the two chains at quarter filling.  
We find simultaneously that there appears an SDW on the 
other chain; the phase of the SDW is shifted by $3\pi/4$ 
from the other SDW.  It might be as if the two SDW states 
coexist.  

Because a sign of $S(4n)$ is opposite to those of 
$S(4n \pm 1)$, the spins on the neighboring two sites 
should be opposite.  The semilog plot in the inset of 
Fig.~\ref{fig6}(b) shows the exponential decay of the 
spin-spin correlations.  The correlation lengths are 
$\xi\sim 4.17$ and $3.57$ for $V_2/t_2=8$ and $10$, 
respectively.  $S(4n-1)$ and $S(4n-3)$, which 
correspond to the spin-spin correlations between two 
$t_2$-chains, decay with the same correlation length 
as $S(4n)$.  These behavior are similar to the case 
of the two-leg Heisenberg (and half-filled Hubbard) 
ladder system.  The correlation length is close to 
that for the two-leg isotropic Heisenberg system 
$\xi\sim 3.19$.~\cite{Whi94}  This means the effective 
hopping between the next-nearest-neighbor sites along 
a $t_2$-chain, i.e., sites $i$ and $i+4$, is of the 
order of $t_1$.  Furthermore, we have found that 
$S(l)$ decays exponentially even in metallic regime 
near the $4k_F$-CDW insulating phase.

\begin{figure}
\includegraphics[width=6cm]{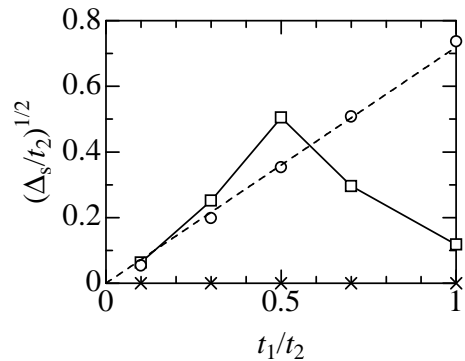}
\caption{\label{fig7} 
Spin gap $\Delta_s/t_2$ as a function of $t_1/t_2$ 
for $V_2/t_2=0$ (crosses), $5$ (squares), and $8$ 
(open circles) at $V_1/t_2=5$ in the thermodynamic 
limit $L \rightarrow \infty$.  
The dashed line is $\Delta_s = 0.5 (t_1/t_2)^2$
}
\end{figure}

From the exponential decay of the spin-spin correlations, 
we expect the system to be in a gapped spin-liquid state.  
In order to confirm it, we have calculated the spin gap 
defined by 
\begin{equation}
\Delta_s(L)=E_0(N_\uparrow+1,N_\downarrow-1,L)
-E_0 (N_\uparrow-1,N_\downarrow+1,L)
\label{spingap}
\end{equation}
for an $L$-site chain.  In Fig.~\ref{fig7}, we show the 
extrapolated values $\Delta_s$ 
$(=\Delta_s(L\rightarrow\infty))$ at $V_1/t_2=5$ for 
$V_2/t_2=0$ ($4k_F$-CDW phase), $5$ (metallic phase), 
and $8$ ($2k_F$-CDW phase).  In the $4k_F$-CDW phase, 
we find that $\Delta_s$ is always zero, which is quite 
natural since the system can be presumed to be a 1D 
Heisenberg model as mentioned above.  
In the $2k_F$-CDW phase, we find that the spin gap is 
always finite except at $t_1=0$ and the size of the 
gap can be scaled with $t_1^2$, i.e., 
\begin{equation}
\frac{\Delta_s}{t_2}\simeq \frac{1}{2}
\left(\frac{t_1}{t_2}\right)^2, 
\label{t_1^2scale}
\end{equation}
as expected from the assumption that the neighboring two 
sites couple with the exchange interaction $J$.  
Because the spin gap for a two-leg isotropic Heisenberg 
ladder is $\Delta_s \sim J/2$,~\cite{Whi94} we can 
estimate the effective exchange interaction as 
\begin{equation}
\frac{J}{t_2} \sim \left(\frac{t_1}{t_2}\right)^2.
\label{effexint}
\end{equation}
As for the metallic phase, $\Delta_s$ goes through a 
maximum at a finite value of $t_1/t_2$ because an 
increase of $t_1/t_2$ leads to an increase in the 
effective exchange coupling, as well as a decrease 
in the $2k_F$-CDW fluctuations.  This behavior reminds 
us of the spin gap of the two-leg CuO ladder materials 
where the gap behaves similarly as a function of the 
energy difference between the O and Cu 
sites.\cite{Nishimoto02}

\subsection{Luttinger parameter}

\begin{figure}
\includegraphics[width=6cm]{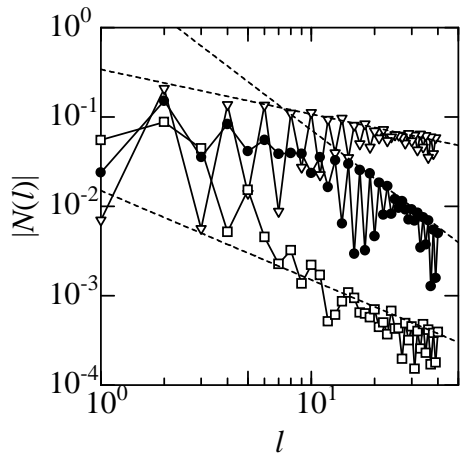} 
\caption{\label{fig8} Density-density correlation 
function $N(l)$ defined in Eq.~(\ref{denscorr}) 
plotted on a log-log scale.  
The results are for $V_2/t_2=3$ (filled triangles), 
$4$ (open squares), and $6$ (filled circles) at 
$V_1/t_2=5$, with $L=128$.  
The distance $l$ is measured along the 1D chain 
of Fig.~\ref{fig1}(c) in units of the interval 
between two sites.}
\end{figure}

Let us calculate the Luttinger parameter $K_\rho$ which 
enables one to know competitions among correlations in 
the metallic phase.  The Hamiltonian 
Eq.~(\ref{hamiltonian}) has the energy dispersion 
relation
\begin{equation}
\varepsilon (k) =-2t_1 \cos k - 2t_2 \cos 2k.
\label{dispersion}
\end{equation}
in the absence of interaction.  If $t_1\ll t_2$ and 
$n=\frac{1}{2}$, the Fermi surface has four Fermi 
points $\pm k_{F_1}$ and $\pm k_{F_2}$, and even for 
finite interactions, one may expect that the system 
at low energies should behave as a two-band 
model.~\cite{Fab96}  In the two-band model, it is 
known that only $4k_F$-CDW and {\it d}-wave 
superconducting (SCd) correlations decay as power 
laws, and all other correlations decay exponentially; 
i.e., either SCd correlation or $4k_F$-CDW correlation 
dominates for both weak and strong interactions.~\cite{Sch96}  
The $4k_F$-CDW correlation and SCd correlation at large 
distances can be written like 
\begin{equation}
N(l) \approx\cos(4k_Fl)\,l^{-2K_\rho},
\label{cdwdecay}
\end{equation}
and
\begin{equation}
\langle\Delta_{i+l}^\dagger\Delta_i\rangle
\approx l^{-1/2K_\rho},
\label{scddecay}
\end{equation}
with 
\begin{equation}
\Delta_i = \frac{1}{\sqrt{2}}
\left(c_{i\uparrow}c_{i+1\downarrow}
-c_{i\downarrow}c_{i+1\uparrow}\right),
\label{scoderparam}
\end{equation} 
respectively.  

In Fig.~\ref{fig8}, we show the density-density 
correlation function $N(l)$ defined in 
Eq.~(\ref{denscorr}) on a log-log scale in order 
to estimate the exponent of the decay.  At large 
distances ($l\gtrsim 10$), the envelope of the 
correlation is consistent with a straight line on 
the log-log scale and therefore one may say that 
it decays as a power law.  Fitting the decay to 
Eq.~(\ref{cdwdecay}) allows us to determine the 
value of $K_\rho$ although the fitting includes 
some errors due to Friedel oscillations in the data.  
The values obtained are given in Table~\ref{table1}, 
which would be of sufficient accuracy when we 
consider available experimental results.  

\begin{table}
\caption{Calculated values of the Luttinger parameter 
$K_\rho$ and charge velocity $v_c$ at $V_1/t_2=5$.}
\label{table1}
\begin{ruledtabular}
\begin{tabular}{ccccccc}
&\multicolumn{1}{c}{$V_2/t_2$}&
\multicolumn{2}{r}{$K_\rho$}& 
\multicolumn{2}{r}{$v_c$}&\\
\colrule
& 2 & &  0.55 & & 0.295 &\\
& 3 & &  1.25  & & 0.738 &\\
& 4 & &  0.88  & & 0.401 &\\
& 5 & &  0.27 & & $< 0.01$ &\\
\end{tabular}
\end{ruledtabular}
\end{table}

At $V_2/t_2\simeq 2.5$, $N(l)$ decays most rapidly 
and $K_\rho$ is the largest.  This means that the 
long-range interactions are effectively weakened 
since $V_1$ and $V_2$ are canceled with each other.  
Such a situation seems to be realized around 
$V_2/t_2=V_1/(2t_2)$.  Generally, long-range Coulomb 
repulsions suppress the value of $K_\rho$.  In our 
results, $K_\rho$ decreases when the values of $V_1$ 
and $V_2$ deviate from the relation $V_1=2V_2$, whereby 
the effective interaction strength increases.  
Now we note that the $2k_F$-CDW transition occurs 
when $K_\rho=1/4$, which may be understood if we 
suppose that two independent $t_2$-chains at quarter 
filling undergo a `$4k_F$'-CDW transition.  
On the other hand, the $4k_F$-CDW transition occurs 
when $K_\rho=1/2$.  This may be understood if we 
assume that the system consists of a half-filled 
chain and an empty chain in the $4k_F$-CDW phase.

We note that the SCd correlations decay more rapidly 
than the $4k_F$-CDW correlations in the whole metallic 
phase even if $K_\rho>1$.  The inverse relationship 
of exponents predicted by a bosonization 
picture~\cite{Bal96,Nag94} seems not to be preserved.  
The same situation was also seen for the two-leg 
Hubbard ladder,~\cite{Noa96} where it is known that 
enhancement of the SCd correlation is not adequately 
observed due to the finite-size effect, i.e., the 
widely separated energy levels near the Fermi energy.  

The slopes at $1/L \rightarrow 0$ in Fig.~\ref{fig2}(a) 
yield $\pi v_c/(2K_\rho)$, from which we can estimate 
the charge velocity $v_c$.  We find that $v_c$ goes 
through a maximum around $V_2/t_2=2.5$ where the 
electrons can move almost freely despite the effect 
of double occupancy on a site.  We note that the 
calculated very small charge velocities near the $2k_F$- 
and $4k_F$-CDW states (see Table~\ref{table1}) may 
suggest heavy-Fermionic features, for which the 
geometrical frustration may play a crucial 
role.~\cite{Shi88,Kon97,Fuj01}

\subsection{Phase diagram}

In Fig.~\ref{fig3} we show the ground-state phase 
diagram of the system in the parameter space of the 
intersite Coulomb interactions $V_1/t_2$ and $V_2/t_2$.  
The phases are distinguished by the presence of a gap 
for spin and/or charge excitations.  Naturally, a 
CDW state is expected if either $V_1/t_2$ or $V_2/t_2$ 
is large.  The ground state of the $2k_F$-CDW 
($4k_F$-CDW) phase is a CO state with (without) spin 
gap and always insulating.  
Then, if both $V_1/t_2$ and $V_2/t_2$ are sufficiently 
small or the effective long-range Coulomb interaction 
is weak due to frustration between two interactions, 
the system may be in a metallic phase.  The metallic 
phases can be divided into two regions.  One is the 
spin-gapped-liquid (SGL) phase where there is a finite 
spin gap but no charge gap.  Here, the system is under 
the influence of $2k_F$-CDW transition and the $2k_F$-CDW 
fluctuation is dominant.  Another is the TLL phase where 
neither spin gap nor charge gap is present and the 
$4k_F$-CDW fluctuation is dominant.

For a lucid discussion, let us now assume the case 
$U/t_2\rightarrow\infty$ and consider some limiting 
cases.  When $V_1/t_2=0$, the $2k_F$-CDW transition 
occurs at $V_2/t_2\simeq 2.3$.  If the system can be 
regarded as two separate $t_2$-chains due to very small 
$t_1/t_2$, this transition would be equivalent to the 
$4k_F$-CDW transition of the quarter-filled spinless 
Hubbard chain with the nearest-neighbor hopping $t_2$ 
and the next-nearest-neighbor Coulomb repulsion $V_2$, 
where the critical value of $V_2/t_2$ is expected to 
be 2, which is consistent with our results.  

When $V_2/t_2=0$, the $4k_F$-CDW transition occurs around 
$V_2/t_2\simeq 2$.  We can estimate this value from 
comparison between the energies of the metallic and 
insulating solutions.  First, the metallic solution may 
be obtained as follows.  Consider the system consisting 
of two independent quarter-filled $t_2$-chains.  Then, 
if we assume the Coulomb repulsion between the 
$t_2$-chains, the total ground-state energy may roughly 
be evaluated as 
\begin{equation}
2E_0^{t_2}(n=\frac{1}{2})+\frac{V_1}{4}L.  
\label{metal-4kf}
\end{equation}
Here $E_0^{t_2}(n=\frac{1}{2})$ is the ground-state 
energy of a spinless-Fermion model defined on a single 
chain at quarter filling with the nearest-neighbor 
hopping $t_2$, and $V_1/4$ is the average Coulomb 
interaction.  Next, the insulating solution may readily 
be obtained because it corresponds to the $4k_F$-CDW 
state, i.e., one of the $t_2$-chains is at half filling 
and another $t_2$-chain is empty, whereby we obtain 
the ground-state energy to be almost zero.  Hence, 
the condition for emergence of the $4k_F$-CDW state is that 
the value of Eq.~(\ref{metal-4kf}) is larger than zero.  
If we use an estimation 
$E_0^{t_2}(n=\frac{1}{2})\simeq -0.25t_2L$ 
made from the Lieb-Wu equations,\cite{Lieb} the critical 
interaction strength can be obtained as $V_1/t_2 \simeq 2$.  
We should note that, because we have used a small but 
finite value of $t_1/t_2$, the actual critical values 
of the interactions calculated are rather larger than 
the above estimation.  

Next, we turn to the case where both $V_1/t_2$ and 
$V_2/t_2$ are nonzero.  Naturally, the system is 
metallic for small $V_1$ and $V_2$ 
($\lesssim {\cal O}(t_2)$), and the metallic phase 
becomes narrower as the long-range Coulomb interactions 
increase along the line $V_2/t_2=V_1/(2t_2)$.  For very 
large values of $V_1$ and $V_2$, the boundary between 
the $2k_F$-CDW (the $4k_F$-CDW) and the SGL 
(TLL) phases becomes parallel to the line 
$V_2/t_2=V_1/(2t_2)$, and the distance between the 
two phases remains to be of the order of $t_2$ ($t_1$) 
even in the limit of $V_1\rightarrow\infty$ and 
$V_2\rightarrow\infty$.  Even if there is no 
hybridization, i.e., $t_1=t_2=0$, the charge mode is 
gapless on the line $V_2/t_2=V_1/(2t_2)$.  This means 
that the effects of $V_1$ and $V_2$ are precisely 
cancelled out on that line.  Now, if finite values 
of $t_1$ and/or $t_2$ are introduced, the energy gain 
due to moving holes expands the width of the line.  
We point out that the width of metallic regime depends 
on the difference $\delta V$ between two repulsive 
interactions $V_1$ and $V_2$ ($\delta V=V_1-V_2$), 
rather than the values $V_1$ and $V_2$ themselves.  
To check this notion, we calculate the critical strength 
of $V_2/t_2$ for very large value of $V_1/t_2$.  
The critical interaction strengths for the $4k_F$-CDW 
and $2k_F$-CDW transitions are obtained respectively 
as follows: $V_2/t_2\approx 249.5$ and $253.5$ for 
$V_1/t_2=500$, and $V_2/t_2 \approx 499.5$ and $503.5$, 
where we assume $V_1/t_2=1000$ and $U/t_2=4000$.  
If $t_1=0$ ($t_2=0$), the boundary of the $2k_F$-CDW 
(the $4k_F$-CDW) phase should be asymptotic to the 
line $V_2/t_2=V_1/(2t_2)$.  

\section{SUMMARY AND DISCUSSION}

\begin{figure}
\includegraphics[width=6cm]{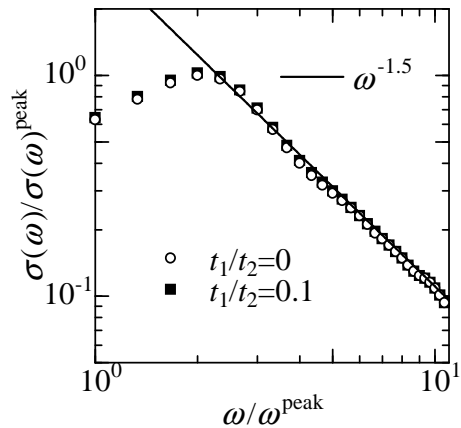} 
\caption{\label{fig9} 
Normalized optical conductivity on a log-log scale.  
Parameter values used for calculation are: 
$U/t_2=20$, $V_1/t_2=2$, $V_2/t_2=5$, $L=32$, 
and $n=0.46875$.
Finite broadening width $\eta=0.4t_2$ is introduced.  
The solid line shows a fitting to the form 
$\sigma(\omega) \propto \omega^{-\gamma}$.  
$\gamma \approx 1.5$ is found.
}
\end{figure}

Using the DMRG method, we have studied the electronic 
states of a two-chain Hubbard model at quarter filling 
that models the low-energy electronic states of the 
CuO double chains of PrBa$_2$Cu$_4$O$_8$.  
We have shown that the model has the CO phases with 
the stripe-type and in-line--type patterns in the 
parameter space and between the two CO phases there 
appears a wide region of vanishing charge gap, which 
is due to geometrical frustration of the long-range 
Coulomb interactions.  
The phase diagram of our model consists of four phases 
characteristic of 
(i) the stripe-type CO (or $2k_F$-CDW), 
(ii) in-line--type CO (or $4k_F$-CDW), 
(iii) Tomonaga-Luttinger liquid, and 
(iv) spin-gapped liquid.  
For the liquid phases, we have suggested possible 
emergence of a heavy-Fermion-like behavior, which 
can occur near the region of the metal-insulator 
phase boundary.  

Finally, let us discuss consistency between our 
numerical results and available experimental data on 
PrBa$_2$Cu$_4$O$_8$.  Comparable experimental features 
are the appearance of the TLL behaviors and estimation 
of the value $K_\rho\sim 0.24$.\cite{Tak00,Miz00,Miz02}  
We can explain such a small value of $K_\rho$ only by 
assuming that the system is in the $2k_F$-CDW state or 
in the SGL state near the $2k_F$-CDW phase boundary.  
In these two phases of our phase diagram, however, a 
small but finite spin gap clearly exists except at 
$t_1=0$.  We have also confirmed that the spin gap 
remains open even if the system is doped slightly 
away from quarter filling.  
Thus, the phase is not TLL unless $t_1=0$, which is 
apparently not consistent with experiment.  
One possibility for reconciliation may be that the 
CuO double chain system has a very small value of 
$t_1$, so that experiments performed with finite 
resolution cannot detect the effects of the spin gap.  
Another possibility is that the optical and ARPES 
spectra observed in relatively high energies are 
insensitive to the opening of the spin gap.  
To demonstrate this, we have calculated the optical 
conductivity of our model with the dynamical DMRG 
method~\cite{Eric02} and confirmed that the shape of 
the higher energy part ($\hbar \omega >\Delta_c$) of 
the conductivity does not depend on the existence of 
small values of $t_1$.  In Fig.~\ref{fig9}, we show 
an example of the $t_1$-dependence of the optical 
conductivity calculated for a possible set of parameters.  
We find that the spectra for $t_1/t_2=0$ and $0.1$ are 
almost the same and the power-law decay can be seen at 
$\hbar\omega>\Delta_c$.  From a fit of the results to 
the form $\sigma(\omega)\propto\omega^{-\gamma}$, we 
estimate the value $\gamma\sim 1.5$, which corresponds to 
$K_\rho\sim 0.22$.  This value is very close to the value 
obtained experimentally.  Here we point out that the 
long-range CDW ground state cannot reproduce the 
experimental spectrum but a small hole doping of only 
a few \% changes the shape of spectrum largely leading 
to the power-law decay of the spectra at high energies.  
We therefore suggest that the actual material is located 
in the SGL phase near the $2k_F$-CDW phase or in the 
$2k_F$-CDW phase with slight doping.  We also want to 
suggest that, if the NMR Knight-shift measurement is 
made on the spin degrees of freedom of the double-chain 
Cu(1) site of PrBa$_2$Cu$_4$O$_8$, one should be able to 
detect anomalous behaviors associated with the spin-singlet 
formation due to the stripe-type fluctuation of charge 
carriers.  

\acknowledgments
We would like to thank E. Jeckelmann for the use of the 
DMRG code originally written by him and useful discussions 
on the numerical techniques.  We also would like to 
thank S. Fujiyama, H. Ikuta, Y. Itoh, and K. Takenaka 
for enlightening discussion on the experimental aspects.  
This work was supported in part by Grants-in-Aid for 
Scientific Research (Nos.~11640335 and 12046216) from 
the Ministry of Education, Culture, Sports, Science, 
and Technology of Japan.  
A part of computations was carried out at the computer 
centers of the Institute for Molecular Science, Okazaki, 
and the Institute for Solid State Physics, University of 
Tokyo.

\end{document}